\documentclass[smallextended]{svjour3}
\smartqed  

\usepackage{amscd,amssymb,amsmath}
\usepackage{graphicx,verbatim,url}
\usepackage{color,slashbox,float,subfloat}
\usepackage{booktabs,tabularx,multirow,array}
\usepackage{caption,eurosym}
\usepackage[round]{natbib}
\usepackage[misc,geometry]{ifsym}
\usepackage[displaymath, mathlines]{lineno}

\newcommand{\var}{\text{var}}

\begin{document}

\title{Multiple changepoint detection for periodic autoregressive models with an application to river flow analysis}

\titlerunning{CHANGEPOINT DETECTION PAR MODELS}        

\author{Domenico Cucina \and Manuel Rizzo \and Eugen Ursu
}

\authorrunning{D. Cucina \and M. Rizzo \and E. Ursu} 

\institute{D. Cucina \at Department of Economics and Statistics, University of Salerno, Via Giovanni Paolo II, 132, 84084 Fisciano, Italy\\
 \and M. Rizzo\;(\Letter) \at Department of Statistical Sciences, Sapienza University of Rome, Piazzale Aldo Moro, 5, 00100 Rome, Italy\\
 \email{manuel.rizzo@uniroma1.it} \\
 \and E. Ursu \at
              GREThA UMR-CNRS 5113, Universit\'e de Bordeaux, Avenue L\'eon Duguit, 33608 Pessac CEDEX, France \\          
}

\date{Received: date / Accepted: date}

\maketitle

\begin{abstract}
In river flow analysis and forecasting there are some key elements to consider in order to obtain reliable results. For example, seasonality is often accounted for in statistical models because climatic oscillations occurring every year have an obvious impact on river flow. Further sources of alteration could be caused by changes in reservoir management, instrumentation or even unexpected shifts in climatic conditions. When these changes are ignored the statistical results can be strongly misleading. This paper develops an automatic procedure to estimate number and locations of changepoints in Periodic AutoRegressive models.
These latter have been extensively used for modelling seasonality in hydrology, climatology, economics and electrical engineering, but there are very few papers devoted also to changepoints detection, moreover being limited to changes in mean or variance. \pagebreak In our proposal we allow the model structure as a whole to change, and estimation is performed by optimizing an objective function derived from the Information Criterion using Genetic Algorithms. The proposed methodology is brought out through the example of three river flows, for which we built models with possible changepoints and evaluated their forecasting accuracy by means of Root Mean Square Error, Mean Absolute Error and Mean Absolute Percentage Error. The last years of data sets have been omitted from the selection and estimation procedure and were then used to forecast. Comparisons with literature on river flow forecasting confirms the efficiency of our proposal.
\end{abstract}

\section{Introduction}
\noindent Discontinuities are often introduced into climatic or hydrologic time series as a result of anthropogenic impacts or changes in instrumentation or location. Further plausible reasons are modifications in reservoir system management or new water pricing.
As defined by~\citet{LL12} a changepoint is "a time where the structural pattern of a time series first shifts". In many cases, changepoints are located at known times (dam construction, measuring instrument change) and it is easy to take into account their effects. When changepoints are located at unknown times and their features are ignored, the time series estimation can be misleading~\citep{LL07, LWLRGF07}. In fact, an undetected changepoint can lead to: misinterpretation of the model, biased estimates and less accurate forecasting~\citep{Hansen01}. Sometimes, even an abrupt shift in climatic conditions does not have a significant impact on hydrologic time series~\citep{BCRCSAS15}. Taking all these into account, changepoint detection becomes a demanding job especially if its identification is required soon after occurrence (e.g. flood predictions).

For the past four decades several techniques have been employed for changepoint detection. As far as hydrological applications are concerned, many authors have considered the problem of detecting a single changepoint~\citep{Cobb78,Buishand84,HM94,RT96}, but very few analyzed more realistic multiple changepoints situations. For example,~\citet{SLX10} proposed a segmented regression with constraint method in order to model both trend analysis and abrupt change detection.~\citet{ZLMWRT15} studied changing trends and regime shifts of streamflow using long term historical records at different hydrological stations in the Yellow River basin.
A Bayesian method is used to investigate the possibility of changes in the mean values and to locate the changepoint in a hydrological time series ~\citep{RT96}. \citet{Kehagias04} and~\citet{KNP06} adopt hidden Markov model, and dynamic programming algorithm, respectively, to locate multiple changepoints in hydrological and environmental time series. A nonparametric estimation of both number and positions of changepoints in multivariate settings is proposed by~\citet{MJ14}. \citet{KS12} use a nonparametric algorithm to solve the problem of changepoint detection by comparing probability distributions over two consecutive intervals. \citet{SLW17} and \citet{GSL17} use, respectively, a dynamic factor model and dynamic programming for studying the segmentation of multivariate time series. In order to assess the dependence of hydrological drought variables,~\citet{SS10} construct a multivariate probability distribution using copulas and estimate its parameters using genetic algorithms. The cumulative sum (CUSUM) approach has proven to be an efficient method to identify the change of annual maximum rainfalls in southern Taiwan~\citep{CPL12}. Several methods (Pettitt, CUSUM, Segment Neighbourhood) for identifying multiple changepoints are presented by~\citet{PG17}; they also show how the presence of autocorrelation may affect these methods.
The problem of modeling a class of nonstationary time series using piecewise Autoregressive processes is considered in~\citet{DLRY06}. A more general procedure for piecewise threshold autoregressive models (TAR) is developed by~\citet{YTL15}. For a recent review of changepoint analysis in time series, see~\citet{AH13}.

Models accounting for seasonality, as the seasonal autoregressive integrated moving average (SARIMA), developed originally by~\citet{BJ70}, have been broadly used in the literature of hydrologic models~\citep{MD05,Du10,WGCZ14}.
Periodic time series models have been introduced because these latter cannot be filtered to achieve second-order stationarity, and this is because the correlation structure of these time series depends on the season~\citep{Ve85a,Ve85b, Mc93}. \cite{MBF17} also showed that the seasonal differencing maintains seasonal correlation structure, whereas periodic term is completely removed by seasonal standardization or by spectral analysis. Using seasonal differenced models can deteriorate the forecasting ability of river flows~\citep{DTK76} which is a main problem for planning and operating of water resources. Also extensions of periodic models have been employed in hydrological applications~\citep{Per17}. General overviews of periodic models and their applications is presented in~\citet{HM94},~\citet{FP04},~\citet{MW06}.

This paper proposes a Periodic AutoRegressive (PAR) model for time series recorded monthly with multiple changepoints at unknown times, which is well-tailored to hydrologic and climatic analysis as the majority of time series observed in these areas are periodic with respect to time. Changepoints detection procedures for such series have been studied by~\citet{LWLRGF07, LLL10} among others. However, PAR models analyzed in these research papers allow mean changes, while the model studied here allows also PAR and trend parameters to switch at each changepoint time. Our results share a number of similarities with their finding allowing at the same time a generalization of results. Moreover, the portmanteau test~\citep{Mc94} designed for diagnosing the accuracy of PAR models is used for each segment. We employ a procedure based on Genetic Algorithms (GAs) to detect changepoints and estimate resulting PAR models, allowing also to identify more parsimonious subset selections. These kind of methods are well suited for complex global optimization, as they have been widely applied to hardly tractable identification and estimation problems~\citep{SS10,UP15}. General proposals of changepoint detection for time series by means of GAs can be found in~\citet{JK13, DFHW17} among others.

Our method is employed to analyze the average monthly flows of three rivers: the Garonne (France), and two Canadian rivers, South Saskatchewan (measured at Saskatoon) and Saugeen (measured at Walkerton). Forecasting ability of resulting models is then evaluated by standard performance measures. We found that models with changepoints (possibly due to both climatic reasons and human intervention) outperforms standard PAR models in terms of prediction accuracy. This could led to improvements in operating the reservoir system, which can result in saving large amount of money (an overview of optimization methods used in reservoir operation is presented in~\citet{FEJ13}). The rest of the article is organized as follows: Section \ref{sec:method} describes proposed methodology for estimation and forecasting; data and discussion on the results are included in Section \ref{sec:results}; comments close the paper in Section \ref{sec:concl}.

\section{Methodology}
\label{sec:method}

\subsection{Model description and estimation}
We consider the problem of modeling a non-stationary time series by segmenting the series into several PAR processes linked at different changepoints. The period of series is $s$ and is assumed to be known. Observation in season $k$ of year $n$ is denoted by $X_{(n-1)s+k}$, with $n = 1,2,\ldots ,N$ and $k= 1,\ldots, s$.

There are $M$ different segments, each of which contains an integer number of years, and $\tau_{j-1}$ denotes the first year of segment $j=1,2,\ldots,M$. The first segment includes years from $\tau_0 = 1$ to $\tau_1-1$, second segment contains years from $\tau_1$ to $\tau_2-1$, third segment contains years from $\tau_2$ to $\tau_3-1$, and so on. If we indicate with $m=M-1$ the number of changepoint times, the segment structure is defined as follows:
\[1\equiv \tau_0<\tau_1<\ldots<\tau_m< \tau_{M}\equiv N+1.\]

In order to ensure reasonable estimates, it is required that each segment contains at least a minimum number $mrl$ of years, therefore $\tau_j \geq \tau_{j-1} + \hspace{2pt} mrl$ for any segment $j$. We let $R^j = \{\tau_{j-1},\tau_{j-1}+1,\ldots,\tau_j-1 \},j=1,2,\ldots,M$, so that if year $n$ belong to set $R^j$ then time $(n-1)s+k$ is in segment $j$. For the sake of simplicity we assume that the total number of observations $T$ is a multiple of $s$.

The model driving our work is given by:
\begin{equation}
\label{par_breaks}
X_{(n-1)s+k} = a^j + b^j [(n-1)s+k] + W_{(n-1)s+k},\;
\end{equation}
where $n\in R^j,\; j=1,2,\ldots,M, \; 1\leq k\leq s$ and
\[W_{(n-1)s+k}=Y_{(n-1)s+k}+\mu_k^j.\] The process $\{Y_{(n-1)s+k}\}$ is a PAR given by:
\begin{equation}\label{par}
 Y_{(n-1)s+k} = \sum_{i=1}^{p^{j}(k)} \phi^{j}_i(k) Y_{(n-1)s+k-i} + \epsilon_{(n-1)s+k}.
\end{equation}
We assume that trend parameters $a^j$ and $b^j$ depend only on the segment, whereas means $\mu_k^j$ are allowed to change also with seasons. The autoregressive maximum model order at season $k$ in the $j$-th segment is given by $p^j(k)$, so that $\phi^j_{i}(k)$, $i=1,\ldots,p(k)$, represent the PAR coefficients during season $k$ of the $j$-th segment. For simplicity, we assume that maximum autoregressive order $p^j(k)=p,$ $\forall j,k$. In our procedure coefficients $\phi^j_{i}(k)$ will be allowed to be set to zero, in order to get more parsimonious models. This is accomplished by introducing in each segment $j$ a binary vector $\delta^j$ of length $s \times p$, named PAR lags indicator, which specify presence or absence of $\phi^{j}_i(k)$ parameters. The first $p$ digits represent the subset PAR model for period $1$, subsequent $p$ digits are related to period $2$ and so on.

The error process $\epsilon = \{ \epsilon_t, t \in \mathbb{Z} \}$ in equation~(\ref{par}) is a periodic white noise, with $E(\epsilon_{(n-1)s+k}) = 0$ and $\var(\epsilon_{(n-1)s+k}) = \sigma_j^2(k)>0$, $n\in R^j$, $j=1,2,\ldots,M$, $1\leq k \leq s$
Unless otherwise stated we assume that each segment is periodic stationary with period $s$, in the sense that
\[Cov(Y_{n+s},Y_{m+s}) = Cov(Y_n, Y_m),\]
for all integers $n$ and $m$. Periodic stationarity is discussed in~\citet{Gl61}.

The number of changepoints $m$, the changepoints location $\tau_1,\tau_2,\ldots,\tau_m$ and PAR lags indicator $\delta^1,...,\delta^M$ are named as structural parameters. They can take discrete values and their combinations amount to a very large number. Once such parameters are determined, the trend intercepts  $a^j$, the slopes $b^j$, the seasonal means $\mu_k^j$, the AR parameters $\phi_i^j(k)$ and the innovation variances $\sigma_j^2(k)$ (for segment $j$, season $k$ and lag $i$) are analytically estimated. These parameters, assuming that structural parameters are known, can be estimated according to the following steps:
\begin{enumerate}
\item Trend parameter vectors $a =(a^1,...,a^M)$ and $b =(b^1,...,b^M)$  are estimates by Ordinary Least Squares (OLS) method:
\[
\min_{a, b} \sum_{j=1}^{M} \sum_{k=1}^s  \sum_{n \in R^j} \left( X_{(n-1)s+k}-a^j-b^j[(n-1)s+k] \right) ^2,
\]

that leads to detrended data \[\hat W_{(n-1)s+k} = X_{(n-1)s+k} - \hat a^j - \hat b^j[(n-1)s+k], \,\,\, n \in R^j, j=1,...,M, 1 \leq k \leq s\]

\item Seasonal means $\hat \mu_k^j$ are computed on resulting detrended data as follows:
\[
\hat \mu_k^j= \frac{1}{\tau_j-\tau_{j-1}} \sum_{n \in R^j} \hat W_{(n-1)s+k}
\]

and implies: $\hat Y_{(n-1)s+k}= \hat W_{(n-1)s+k} - \hat \mu_k^j$.

\item AR parameters are estimated separately for each segment and season. Each of these specific series $z_k^j$ is selected from $\hat{Y}$ and is incorporated in a design matrix $Z$ of dimensions $(\tau_j-\tau_{j-1}) \times p$, which includes lagged observations. Parameter constraints are specified by a $(p-q) \times p$ matrix $H$, where $q$ is the number of free parameters. These constraints are designated on the basis of PAR lags indicator $\delta^j$ as follows:
\begin{itemize}
	\item For each lag $i$, the element $[p(k-1)+i]$ of $\delta^j$ vector is evaluated
	\item If value is equal to 1 then a row equal to the $i$-th row of $I_p$ identity matrix is added to $H$.
\end{itemize}

Final estimate $\hat \phi^j(k)= (\hat \phi^j_1(k), \ldots,\hat \phi^j_p(k))$  of $\phi^j(k)$ is obtained by constrained optimization, with linear constraint given by $H \phi^j(k)=0$. Explicitly (in matrix form):
\[
\hat \phi^j(k)= \phi^{j,LS} (k)  - (Z'Z)^{-1} H' [ H(Z'Z)^{-1}H']^{-1}H \phi^{j,LS} (k), \\
\]

where $\phi^{j,LS} (k)= (Z'Z)^{-1}Z' z_k^j$ is obtained by OLS estimation. \\

\item Lastly, estimation of innovation variances $\hat{\sigma}_j^2(k)$ is performed for each segment and season on final residuals, considering that each regime has a possibly different sample size. \\

\end{enumerate}

Selection of optimal structural parameters, on the other side, is a complex problem for which no closed form solution is available. As far as it involves the evaluation of a very large number of possible combination, GAs are naturally suitable for this issue.

\subsection{Identification of structural parameters}
\label{subsec_estim}
The GA is a nature-inspired optimization method, often employed when it is required to find an optimal solution from a prohibitively large discrete set.
In a maximization problem it allows to approximate the optimal solution, which maximizes an objective function (named fitness in GA terminology), by a simple procedure. In the generic iteration a population of binary encoded solutions (called chromosomes) is subdued to genetic operators:
the selection randomly chooses chromosomes for the subsequent steps, usually proportionally to their fitness value; by crossover two solutions are allowed to combine together, with a fixed rate $pC$, exchanging part of their values and creating two new individuals; lastly, the mutation step allows each binary value to flip its value from 0 to 1, or vice versa, with a fixed probability $pM$, providing a further exploration of search space. The resulting population replaces the previous, and the flow of generations stops if a certain condition is met, for example a fixed number of generations. It is also possible, adopting the elitist strategy, to maintain the best chromosomes found up to the current iteration, irrespective of operators effect.

Two different genetic approaches have been analyzed in pilot experiments for our structural parameters identification problem. The first method is a hybrid GA (HGA), in which number of changepoints $m$ and changepoint locations  $\tau_1,...,\tau_m$ are encoded in a binary chromosome, while PAR lags indicator are obtained by enumerating all possible subset models, given each segment and season, and returning only the best one. This latter task is computationally feasible when maximum autoregressive order $p$ is small, as far as $2^p$ models must be evaluated for each segment and season. In this way of proceeding optimal values of structural parameters are obtained combining an exact method (exhaustive enumeration) with an approximation procedure (GA). Second algorithm is a simple GA (SGA) in which subset model indicators are included in chromosome. In this case the size of search space and the length of chromosome considerably increase with $p$. As a result of this it was necessary to increase the number of iterations for SGA. Both of these algorithms, however, seemed to be equivalent on exploring the solution space, and for small values of $p$ they very often reach an equivalent solution. As far as in this paper we shall consider values of $p$ up to $3$, because autoregressive procedure must capture the short term dependence, we employed HGA in our analysis. For larger values of $p$ we recommend the use of SGA.

In HGA binary chromosomes encode a candidate segmentation $[m,\tau_1,...,\tau_m]$ as follows: first two bits give number of changepoints $m$ (limited to a maximum of 3 in our study, so that a number of segments up to 4 is allowed); subsequent bit intervals, whose length is custom fixed, produce changepoint times $\tau_1,...,\tau_m$. This part of encoding must ensure following constraints:
\begin{eqnarray*}
mrl+1 &\leq& \tau_1, \hspace{5pt} mrl+\tau_1 \leq \tau_2, \hspace{3pt} ..., \hspace{3pt} mrl+\tau_{m-2} \leq \tau_{m-1}, \\
mrl+\tau_{m-1} &\leq& \tau_m \leq N-mrl-1 ,
\end{eqnarray*}

due to the fact that a minimum number $mrl$ of observations must be contained in each segment. In order to accomplish this the bit intervals encode $m$ real numbers $th_i \in (0,1),\hspace{3pt}i=1,...,m$, constructed to determine percentage of remaining values to be attributed to the corresponding $i$-th segment. In fact, when placing a new changepoint there are some illegal positions, due to above specified constraints: this implies that $mrl$ observations must be left out from both the beginning and the end of considered segment. This strategy depends on candidate number of segments, so changepoints are uniquely identified in four possible ways:

\begin{itemize}
	\item If $m=0$ (one segment) then $\tau_1=N+1$.
	\item If $m=1$ (two segments) then $\tau_1 = mrl+1 + (N-2mrl)\times th_1$ \item If $m=2$ (three segments) then:
	\begin{itemize}
		\item $\tau_1 = mrl+1 + (N-3mrl)\times th_1$
		\item $\tau_2 = mrl+\tau_1 + (N-2mrl-\tau_1+1)\times th_2$
	\end{itemize}	
	
	\item If $m=3$ (four segments) then:
	\begin{itemize}
		\item $\tau_1 = mrl+1 + (N-4mrl)\times th_1$
		\item $\tau_2 = mrl+\tau_1 + (N-3mrl-\tau_1+1)\times th_2$
		\item $\tau_3 = mrl+\tau_2 + (N-2mrl-\tau_2+1)\times th_3$
	\end{itemize}
\end{itemize}

Such an encoding procedure, introduced in \citet{BP12}, allows each possible chromosome to be legal, so there is no computational time wasted on evaluating infeasible solutions.

As far as fitness function measures goodness of solutions in GAs, our model identification problem shall include a term linked to the goodness of fit and a part related to a penalization on number of parameters. Many options are available: we shall consider a criterion inspired by NAIC, introduced by~\citet{To90} for threshold models, given by:
\begin{equation}\label{NAIC}
g= [  \sum_{j=1}^{M} \sum_{k=1}^{s} n_{j,k} \log(\hat{\sigma}_j^2(k)) + IC \sum_{j=1}^{M} \sum_{k=1}^{s} P_{j,k}] / T,
\end{equation}
where $\hat{\sigma}_j^2(k)$ is the model residual variance of series in segment $j$ and season $k$, $n_{j,k}$ is related to the sample size, $P_{j,k}$ is related to the number of parameters, $IC$ is the penalization term. We adopt a scaled exponential transformation of $g$ as fitness function, for maximization purposes: $f=\exp(-g/\beta)$, where $\beta$ is a problem dependent constant which is introduced for suitably scaling fitness.

Before leaving the estimation procedure, we can make a comment. A two-step procedure for identification and estimation of PAR models based on Yule-Walker approach was presented by~\citet{MW06}. We propose an automatic procedure for detecting changepoints, identifying the best model and estimating the parameters for each segment. Our estimation method is based on multiple linear regression. Both estimation techniques are efficient from statistical point of view.

\subsection{Forecasting and performance assessment}
\label{subsec_forec}
The forecasting method employed is the standard one-step-ahead procedure. We shall remove last year from dataset (corresponding to 12 observations) in order to estimate model parameters, and use those data points for evaluating forecasting performance. The logarithm of data is adopted as Box-Cox transformation, performed before fitting the model, as far as it is the most used in monthly river flow analysis and it ensures that model residuals are approximately normal distributed and homoscedastic~\citep{EV92,MG13}.

Forecasting accuracy of proposed models is evaluated with respect to the following measures:

\[
RMSE = \sqrt{\frac{\sum_{i=1}^{12}( y_i - \hat{y}_i )^2}{12}},
\]

\[
MAE = \frac{\sum_{i=1}^{12} | y_i - \hat{y}_i |}{12},
\]

\[
MAPE = \frac{\sum_{i=1}^{12} \frac{| y_i - \hat{y}_i |}{y_i} }{12} \times 100 ,
\]

where $y_1,...,y_{12}$ and $\hat{y}_1,...,\hat{y}_{12}$ are, respectively, true and predicted values of last year observations, on logarithmic scale.
These measures are explicitly defined in~\citet{HK05}. A smaller value indicates a better model performance. The MAE and the RMSE between the proposed model and the observed data are calculated in the same units of the observed data. The MAPE measure is based on percentage errors; these latter have the advantage of being scale-independent, and they are frequently used to compare forecast performance between different data sets. These criteria must be interpreted only as an indication that shows which model performs better, but no statement can be made from this comparison. All measures are calculated using the \textit{hydroGOF} and/or \textit{forecast} packages in R software. For other measures of forecast accuracy we refer to~\citet{HK05, KBB05}.

\subsection{Model validation}
The analysis of the residuals for the estimated model is needed to test its relevance. The stationarity of the residuals allows us to apply the standard 95\% confidence limits (that is $1.96\sqrt{Ns}$).
To test the joint statistical significance of the residual autocorrelations,~\citet{Mc94} proposed a Ljung-Box portmanteau test for PAR models:
\begin{equation}
\label{QL}
Q_L(k) = N \sum_{l=1}^{L} \frac{N}{N-\lfloor (l-k+s)/s \rfloor}
                 r_l(k)^2 ,
\end{equation}
where $k$ designs the period, $r_l(k)$ is the autocorrelation coefficient with lag $l$, $L$ is considered the maximum time lag and where
$\lfloor x \rfloor$ represents the integer part of the real number
$x$.
The test statistic
$Q_L(k)$
follows approximatively
a chi-square distribution
$\chi^2_{L-p(k)}$,
with
$L-p(k)$
degrees of freedom.

\section{Results and discussion}
\label{sec:results}
We shall now study effectiveness of proposed methodology in river flow analysis. Data related to three rivers having different lengths, means of annual flows and located in different regions, will be examined.
They consist of:
\begin{itemize}
\item flows of Garonne river measured at Tonneins, France;
\item flows of South Saskatchewan river measured at Saskatoon, Canada;
\item flows of Saugeen river measured at Walkerton, Canada.
\end{itemize}

Mean, standard deviation and correlation functions of these rivers vary significantly from month to month. Further details will be given in subsequent subsections.

Garonne river has been analyzed in~\citet{UP15}. The South Saskatchewan and Saugeen river flows series are discussed in~\citet{NMH85}, and they are available at
\url{http://www.stats.uwo.ca/faculty/mcleod/epubs/mhsets/readme-mhsets.html}. 
They used nine different models to generate thirty-six one-step-ahead forecasts for the logarithmic flows. In terms of RMSE, PAR models give the best results. We will show that our methodology can improve their results.

Estimation and forecasting will be performed as discussed in subsection \ref{subsec_estim} and \ref{subsec_forec}. As far as choice of genetic operators is concerned we propose standard roulette wheel selection, bit-flip mutation, and a modified single-point crossover: instead of allowing all bits in chromosome to be selected as possible cutting points, we shall consider only bits that subdivide segmentation $[M,\tau_1,...,\tau_m]$. In such a way parameter structures can be naturally inherited by offspring, avoiding the possibility of destroying solutions. Elitist strategy is also employed.

In fitness function, related to NAIC criterion~(\ref{NAIC}), we fixed $IC=2$ so that it resembles the penalization structure of AIC. Several experiments have been conducted considering various combinations of parameters $p$, $mrl$ and $M$, in order to provide a variety of explicative models.
We used the following strategy: conditioning on four possible values of $M$, which include the model with no changepoints and situations with possible structural breaks up to, respectively, $1$, $2$ and $3$, we selected four models for which the best value of fitness function has been observed. Forecasting accuracy of these models, labelled as PAR$_{(M; p; mrl)}\hspace{3pt}(M=0,1,2,3)$,
will be then evaluated.

Computations will be performed by using Matlab and R softwares.

\subsection{Garonne river}
\begin{figure}[ht]
		\centering{}
		\includegraphics[width=0.95\textwidth]{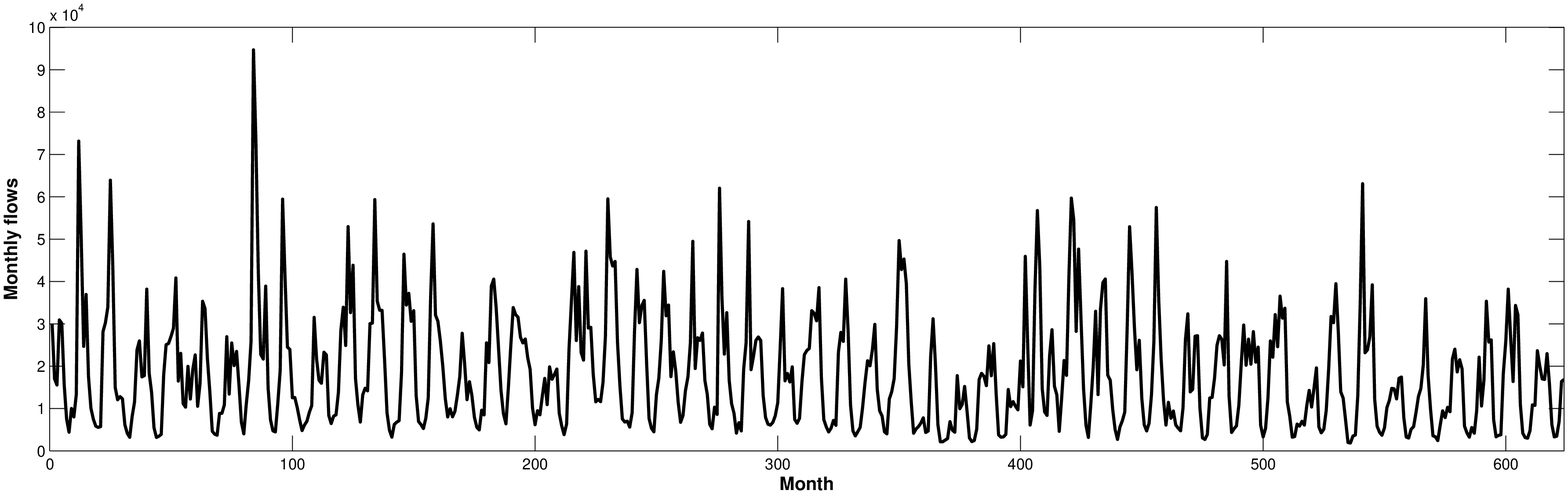}
		\centering{}
		\includegraphics[width=0.95\textwidth]{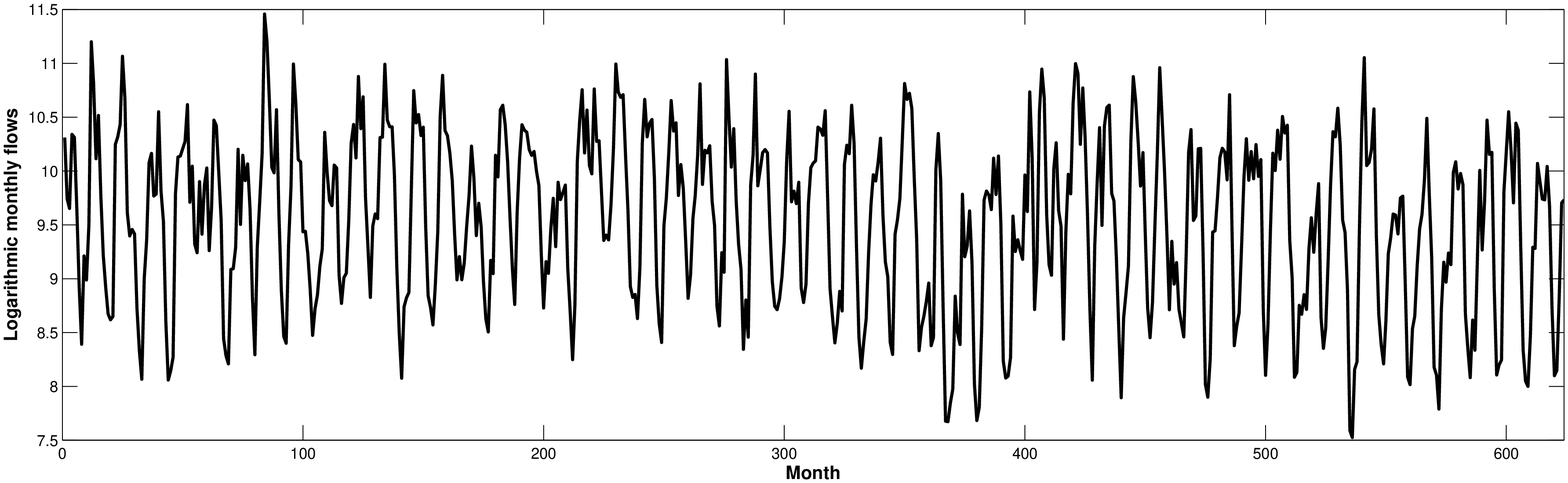}
			
\caption{Monthly flows (up) and logarithmic monthly flows (down) for the Garonne river.}
	\label{fig:garonne_plot}
\end{figure}

The Garonne river which flows through Spain and France is the third largest river in France in terms of flow. Its total length is about 647 km with a catchment area of 51500 km$^2$ at Tonneins. It is the main contributor to the Gironde Estuary which is the major European fluvial-estuarine system.
Flow measures are recorded at the Tonneins gauging station, where there is no tidal effect. Data are obtained from daily discharge measurements in cubic meter per second (m$^{3}$/s) from January 1959 to December 2010 (DIREN-Banque Hydro, French water monitoring). Daily data flows are then transformed in monthly data consisting in flows averaged for one month.
The data and the log transformed data (624 observation or 52 years) are presented in Figure~\ref{fig:garonne_plot}.

Optimal PAR models and forecasting results are reported in Table \ref{table_garonne}. The first proposed model is a PAR without changepoints which is extensively used to perform hydrological forecasting~\citep{NMH85,MW06}. All the other proposed models allow one or more changepoints.

\begin{figure}
\centering
\includegraphics[width=13cm, height=.25\linewidth]{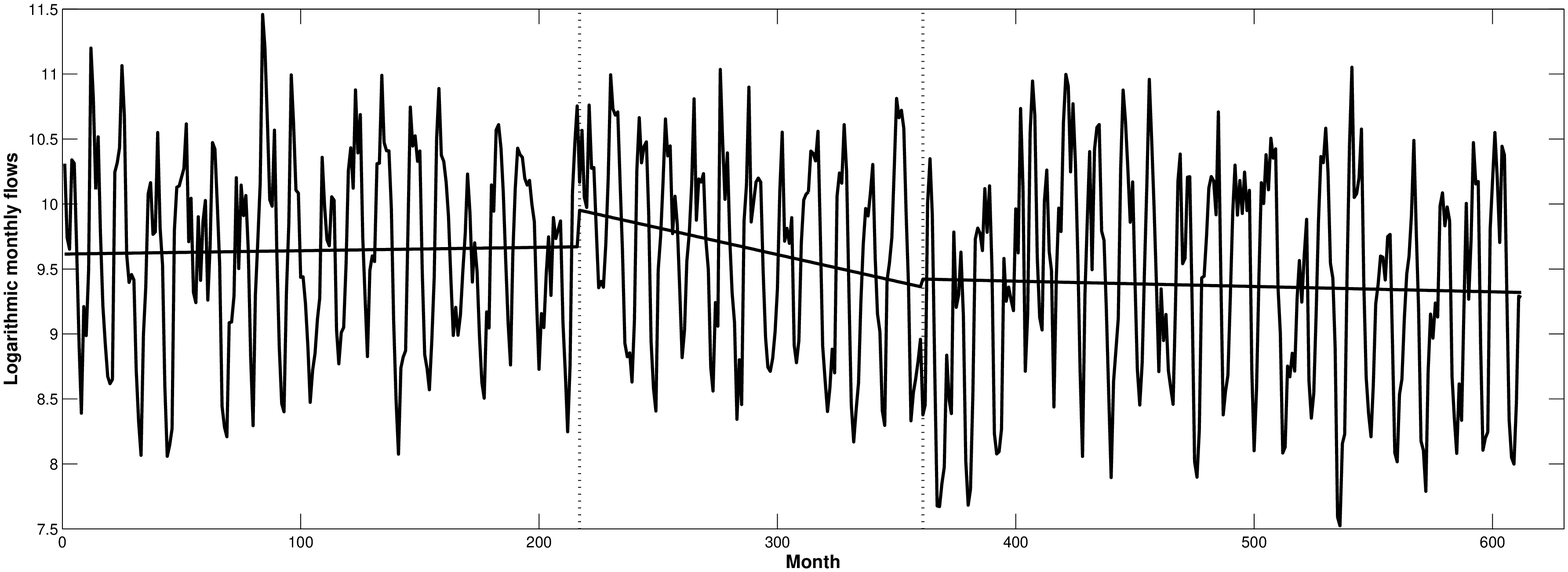}

\caption{Changepoint detected on years 1977 and 1989 for Garonne river}
\label{fig:garonne_changepoint}
\end{figure}

Year 1989 is detected as possible changepoint in almost all configurations. According to~\citet{CVMHNLLB07}, years 1988-1989 seem to be the driest in the 1980-1990 decade. Moreover, the air temperature over Western Europe showed an abrupt shift at the end of the 1980s. For a better understanding of climatic changes and their impact on water resources,~\citet{BCRCSAS15} studied a subset of 119 temperatures, 122 rainfalls and 30 hydrometric stations over the entire France. They detected a shift in annual mean air temperature in 1987-1988 for more than 75\% of the 119 temperature stations. They also detected a shift between 1985 and 1990 for 18 hydrometric stations.

A related series (Dordogne river flows) is available in order to help draw conclusions. Using flow series for Dordogne river from 1959 to 2010, we found exactly the same 2 changepoints (1977 and 1989) as with PAR$_{(2;3;12)}$ model for Garonne river. The Dordogne and Garonne rivers are the main rivers of the Adour-Garonne river basin. The climate in the basin area is under the influence of oceanic conditions in its western parts and we suppose that both rivers experience similar climatic conditions. Changepoint corresponding to 1977 could be linked to local effects. Generally, it is difficult to attribute any of the changepoint time to a specific climatic event or anthropogenic factor.

\begin{table}
\centering
\begin{tabular}{|c|c|c|c|c|c|}
\hline
 & Years of changepoint & $RMSE$ & $MAE$ & $MAPE$ &$Fitness$\\
 \hline
 PAR$_{(0;3)}$ & / & \textbf{0.2476} & 0.2215 & 2.3562 & 1.2022 \\
 \hline
 PAR$_{(1;3;10)}$  & 1989 & 0.2731 & 0.2133 & 2.2665 & 1.2113 \\
 \hline
 PAR$_{(2;3;12)}$  & 1977, 1989 & 0.2725 &\textbf{ 0.2131} & \textbf{2.2643} & 1.2146 \\
\hline
 PAR$_{(3;3;10)}$  & 1970, 1988, 1998 & 0.3149 & 0.2511 & 2.6100 & 1.2241 \\
\hline
\end{tabular}
\caption{Results of evaluation criteria of the logarithmic forecast errors for Garonne}
\label{table_garonne}
\end{table}

In terms of goodness of fit we observe that fitness values are monotonically increasing with number of changepoints; forecasting accuracy, on the other side, is maximum in terms of MAE and MAPE for model with 2 changepoints corresponding to 1977 and 1989, while best value of RMSE is observed for model with no changepoints.

\begin{figure}
\centering
\includegraphics[width=13cm,height=.25\linewidth]{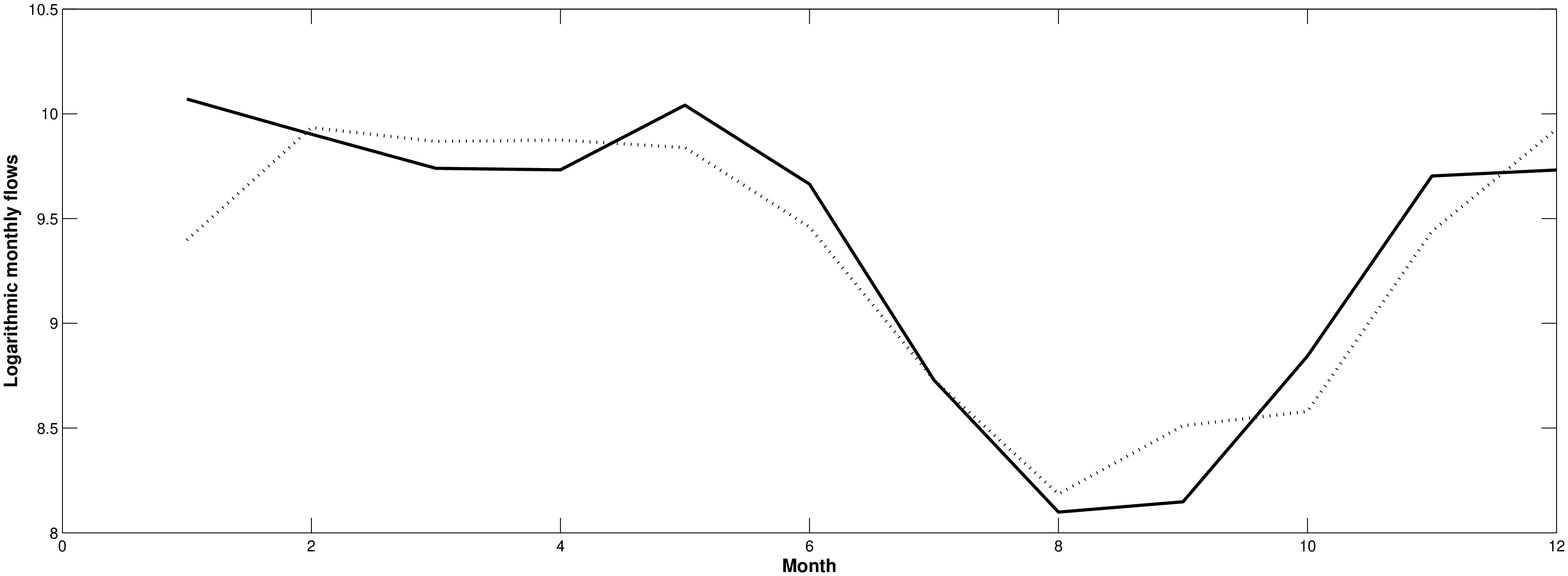}
\caption{Logarithmic flows of Garonne (full line) and one-step best PAR forecasts (dashed line).} \label{fig:log_garonne_forecasts}
\end{figure}

Figure~\ref{fig:garonne_changepoint} graphically displays the segmentation  with 2 changepoints and Figure~\ref{fig:log_garonne_forecasts} shows the logarithmic flows and their forecasts using a PAR model with two changepoints. From latter figure we can see that the RMSE value could be influenced by large error prediction of first observation.

\begin{figure}
\centering
\includegraphics[width=\linewidth, height=5cm] {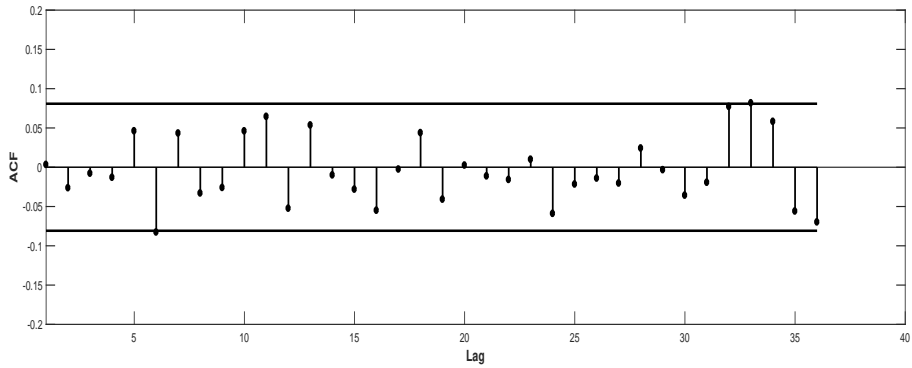}
\caption{Autocorrelation function (ACF) of the residuals of the fitted PAR model with two changepoints to the Garonne flow.} \label{fig:ACF_log_garonne}
\end{figure}

As a diagnostic check, the residual autocorrelations up to lag 36 were computed. Figure~\ref{fig:ACF_log_garonne} shows the autocorrelation function (ACF) for the residuals of PAR model with 2 changepoints. The two full lines indicate lower and upper bounds of the ACF assuming that residuals are white noise. The residuals for PAR model with 2 changepoints indicate no correlation. This graphic checking provides "some evidence" on the adequacy of proposed PAR model. Validity of our model is also evaluated with the portmanteau test given in equation~(\ref{QL}) and results are presented in Table~\ref{pval_garonne}. All P-values in this table suggest that the proposed model is not rejected at the usual 5\% significance level, except for October in the first segment, with an empirical significance level of about 2.5\%. However, this does not strongly point to the model inadequacy as that empirical level is still not too low and, consequently, the possibility of an error of type I exists.

\begin{table}
\centering
\begin{tabular}{|c|c|c|c|}
  \hline
   & 1959-1977 & 1978-189 & 1989-1999 \\
   \hline
  January & 0.4217 & 0.7622 & 0.6915 \\
  February & 0.1701 & 0.2600 & 0.5702 \\
  March & 0.3938 & 0.5050 & 0.1058 \\
  April & 0.9323 & 0.2758 & 0.7183 \\
  May & 0.6557 & 0.0789 & 0.6045 \\
  June & 0.0761 & 0.1109 & 0.5058 \\
  July & 0.6058 & 0.7771 & 0.5916 \\
  August & 0.3658 & 0.6296 & 0.1500 \\
  September & 0.3581 & 0.2201 & 0.8310 \\
  October & 0.0251 & 0.0950 & 0.7068 \\
  November & 0.3318 & 0.1811 & 0.6034 \\
  December & 0.1558 & 0.3261 & 0.5973 \\
  \hline
\end{tabular}
\caption{P-values of the portmanteau test defined in eq.~(\ref{QL}) with $L=15$.}
\label{pval_garonne}
\end{table}

\subsection{South Saskatchewan river}
The South Saskatchewan River originates in the Rocky Mountains and drains an area of about 139,600 km$^2$. It joins the North Saskatchewan river and form Saskatchewan river, which is the fourth longest river in North America.

\begin{figure}[ht]
		\centering{}
		\includegraphics[width=0.95\textwidth]{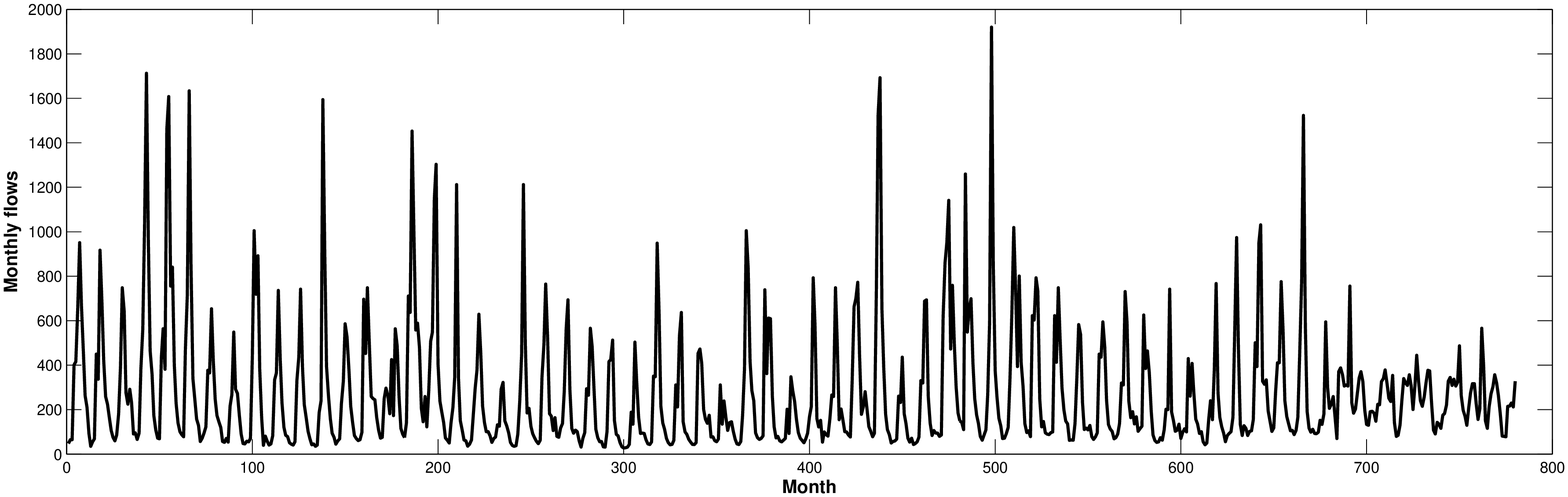}
		\centering{}
		\includegraphics[width=0.95\textwidth]{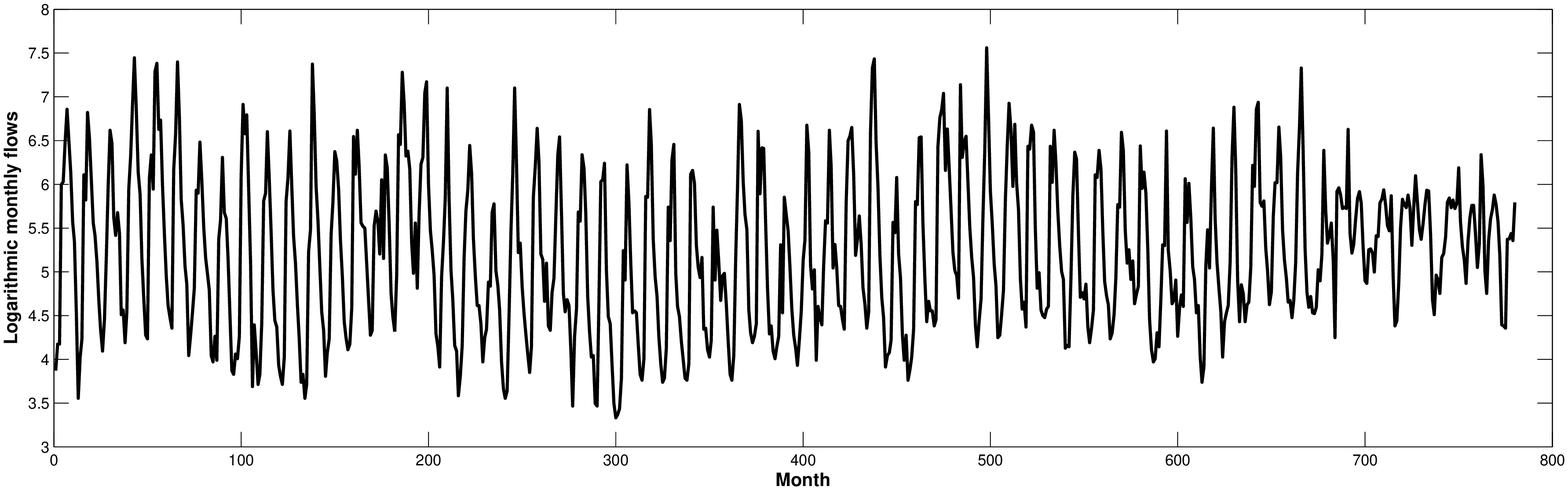}

\caption{Monthly flows (up) and logarithmic monthly flows (down) for the South Saskatchewan river.}
	\label{fig:saskatchewan}
\end{figure}

The time series of mean monthly flows of the Saskatchewan river measured at Saskatoon, Canada from January 1912 to December 1976 include 780 observation (65 years). The monthly series and logarithmic monthly flows can be viewed in Figure~\ref{fig:saskatchewan}.

\citet{NMH85} fitted the transformed data with nine models and they recommended the PAR model. Using a PAR without changepoints we obtain a similar model as in~\citet{NMH85}; the slight differences arise from the fact that our model allows intermediate constraints and we use more data. In addition, our method was applied to estimate a PAR model with at least one changepoint. Optimal PAR models are showed in Table~3.

\begin{figure}
\centering
\includegraphics[width=13cm,height=.25\linewidth]{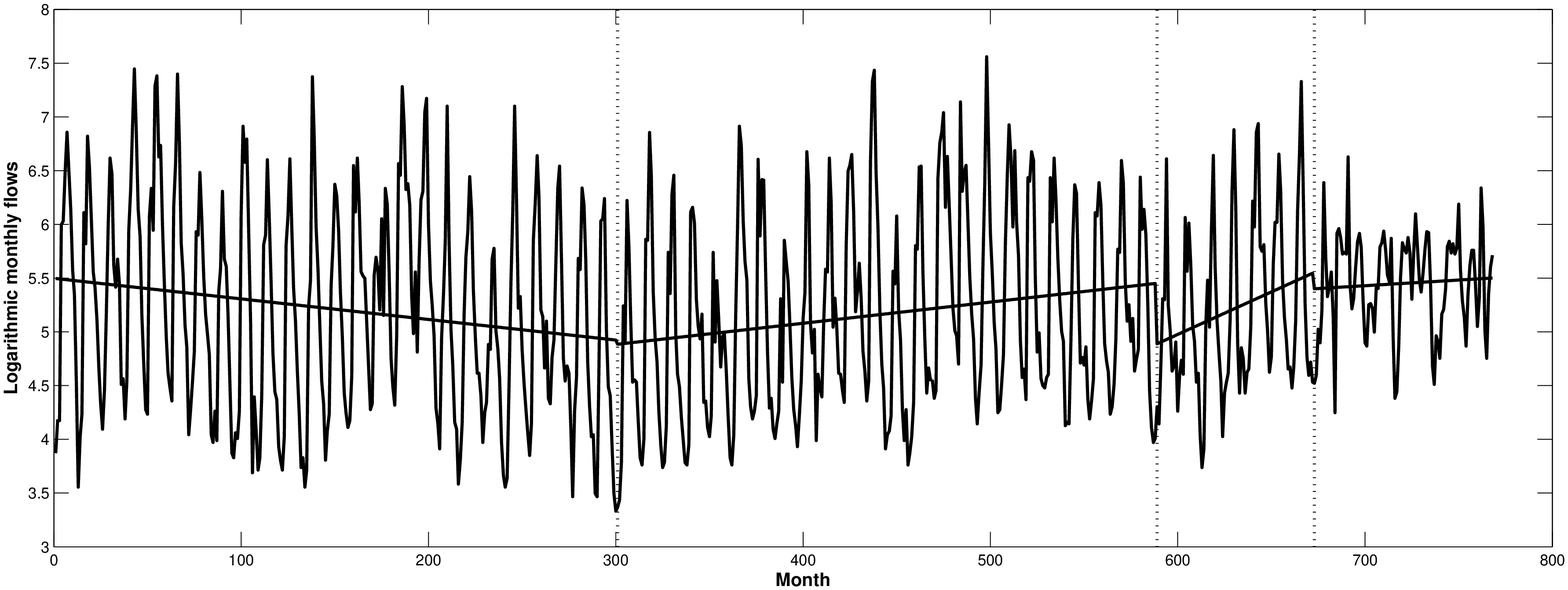}
\caption{Changepoint detected on years 1937, 1961 and 1968 for Saskatchewan river}
\label{fig:Saskatchewan_changepoint}
\end{figure}

In all models with at least one possible changepoint we detect year 1968. January 1969 corresponds to a modification in the reservoir system management: the Gardiner Dam came into full operation upstream from Saskatoon on the South Saskatchewan River. The Gardiner Dam is the third largest embankment dam in Canada and one of the largest embankment dams in the world. The reservoir provided valuable benefits to the community: power generation, recreational benefits, and also magnitude of floods have been lessened, with minimum flows downstream guaranteed throughout the year. Before the creation of Gardiner Dam, snowmelt in conjunction with summer rains led to heavy flooding.

\begin{figure}
\centering
\includegraphics[width=\linewidth, height=5cm]{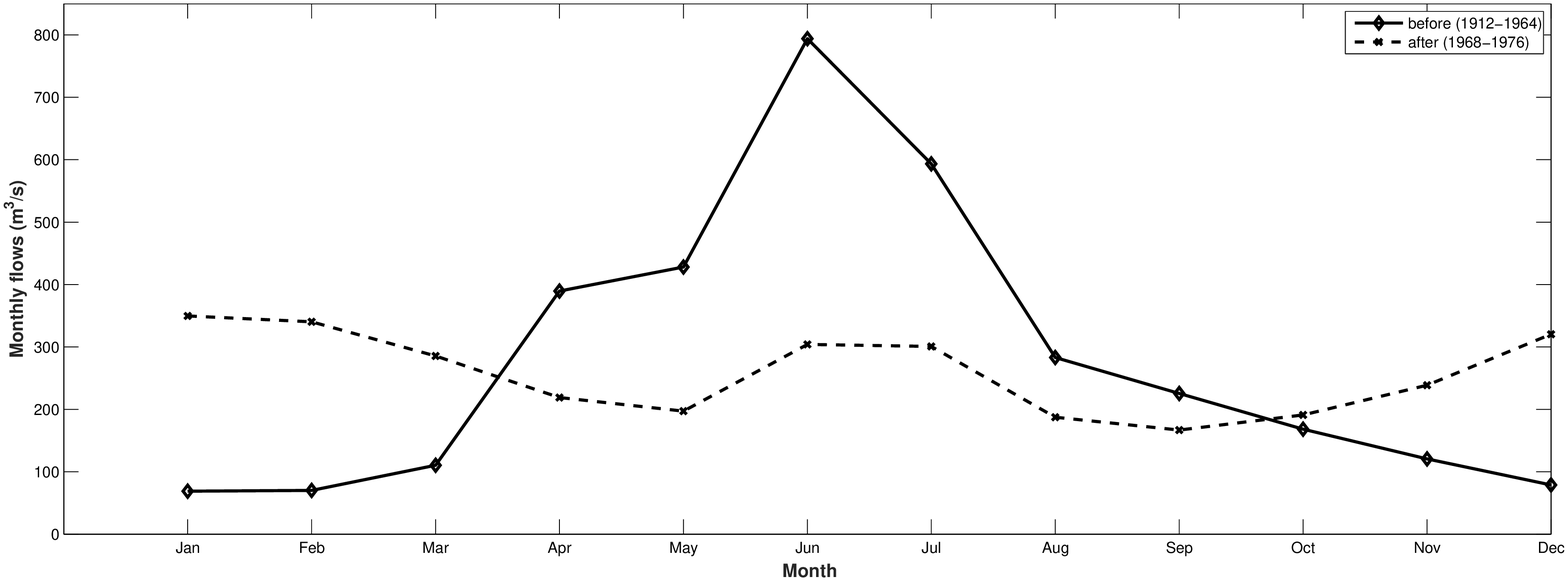}

\caption{Monthly means of the South Saskatchewan river before (full line) and after (dashed line) of the construction of Gardiner dam} \label{fig:saskatchewan_month}
\end{figure}

To model the effect of the operation of the Gardiner dam on the average monthly flows of the South Saskatchewan river~\citet{HMM77} developed an intervention model. To see how the monthly average changes after the intervention, the cumulative sum was calculated and plotted for each month. Using flows measured at Saskatoon from 1942 to 1974, they found that the operation of the Gardiner dam significantly affected the average monthly flows: the average flows increase from November to March and decrease from April to September (Figure~\ref{fig:saskatchewan_month}).

Looking into the history, the changepoint corresponding to 1937 could be linked to the drought conditions on the Saskatchewan prairies during the Dust Bowl years of the 1930s. The drought came in three waves, 1934, 1936, and 1939-1940, but some regions experienced drought conditions for as many as eight years. The changepoint corresponding to 1961 could arise as an effect of the beginning of dam construction in 1964.

\begin{figure}
	\centering
	\includegraphics[width=\linewidth, height=5cm]{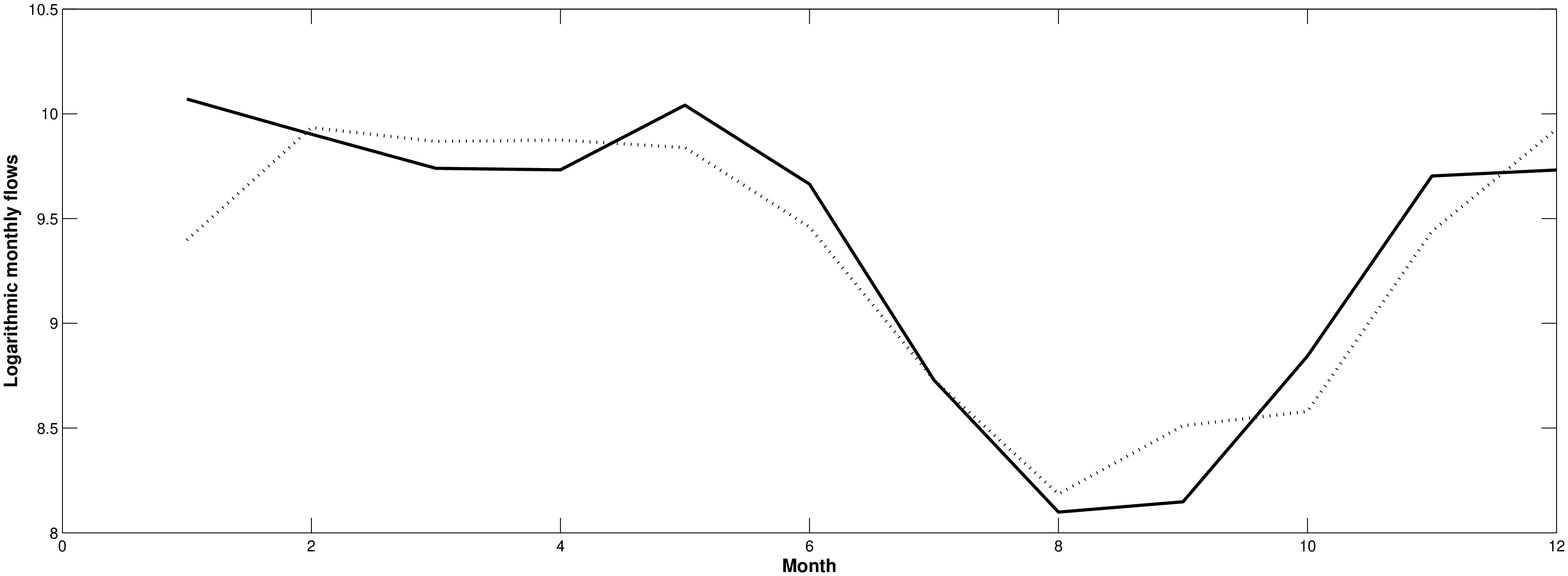}
	\caption{Logarithmic flows of South Saskatcewan (full line) and one-step PAR forecasts (dashed line).} \label{fig:log_saskatchewan_forecasts}
\end{figure}

As for the Garonne river, the performance of models was considered in terms of fitness, as far as goodness of fit is concerned, and RMSE, MAE and MAPE for evaluating accuracy of one step-ahead logarithmic forecasts (Table \ref{table:saskatchewan}). The PAR model with 4 segments (PAR$_{(3;2;7)}$) performs better in terms of both estimation and forecasting in comparison with other models. Figure \ref{fig:Saskatchewan_changepoint} display the logarithmically transformed series with 3 detected changepoints, and  Figure~\ref{fig:log_saskatchewan_forecasts} shows the logarithmic flows and their forecasts using such model.

\begin{table}
\centering
\begin{tabular}{|c|c|c|c|c|c|}
\hline
 & Years of changepoint & $RMSE$ & $MAE$ & $MAPE$ & $Fitness$\\
 \hline
 PAR$_{(0;3)}$ & / &0.66428 & 0.5126 & 10.2718 & 1.2253 \\
 \hline
 PAR$_{(1;3;7)}$  & 1968 & 0.4360 & 0.3167 & 6.3646 & 1.2553\\
 \hline
 PAR$_{(2;3;7)}$  & 1937, 1968 & 0.4348 & 0.3115 & 6.2757 & 1.2643 \\
\hline
 PAR$_{(3;2;7)}$ & 1937, 1961, 1968 & \textbf{0.4275} & \textbf{0.2933} & \textbf{5.8488} & 1.2661\\
 \hline
\end{tabular}
\caption{Results of evaluation criteria of the logarithmic forecast errors for Saskatchewan}
\label{table:saskatchewan}
\end{table}

\begin{figure}
\centering
\includegraphics[width=\linewidth, height=5cm]{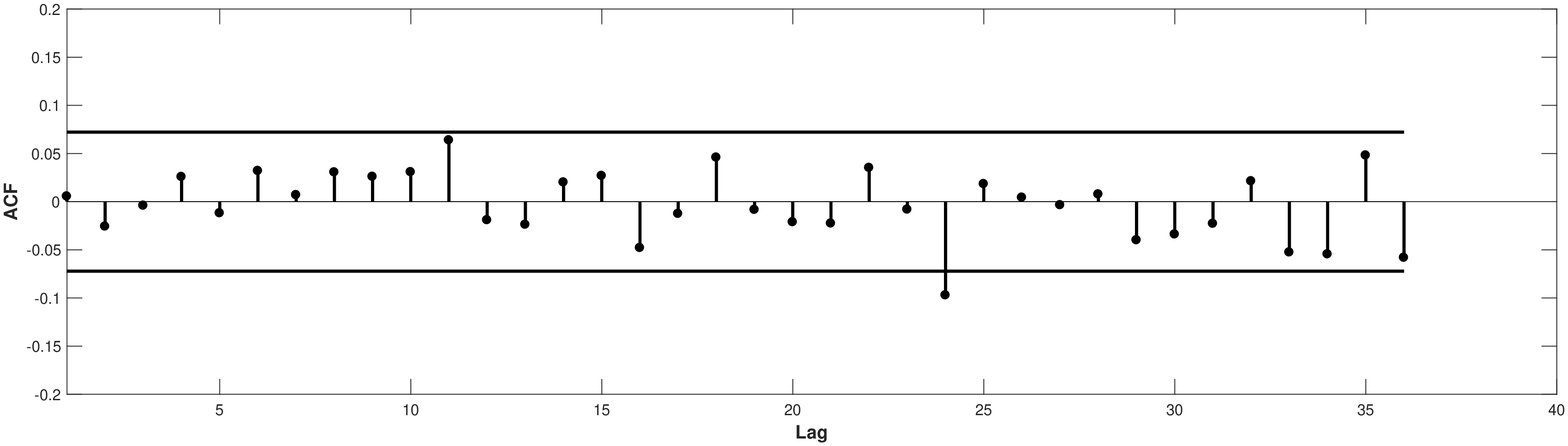}
\caption{Autocorrelation function (ACF) of the residuals of the fitted PAR models with three changepoints to the South Saskatchewan flow.} \label{fig:ACF_log_saskatchewan}
\end{figure}

To check for the whiteness of the residuals the autocorrelations up to lag 36 were computed for the PAR$_{(3;2;7)}$ model. Only one significant autocorrelation was found (Figure~\ref{fig:ACF_log_saskatchewan}). Table~\ref{pval_saskat} shows P-values of portmanteau test: they suggest that the proposed model is not rejected at 5\% significance level, except for August 1961-1968, November 1938-1961 and December 1969-1975 where the P-values are about 1.5\%, which does not strongly suggest model inadequacy. In this situation, PAR model without changepoints seems inappropriate at the 5\% nominal level.

{\color{red}
\begin{table}
\centering
\begin{tabular}{|c|c|c|c|c|}
  \hline
   & 1912-1937 & 1938-1961 & 1961-1968 & 1969-1975\\
   \hline
  January  & 0.9536  &  0.6893  &  0.5412  &  0.0423\\
  February & 0.3981  &  0.5351  &  0.3869  &  0.2332\\
  March & 0.8564  &  0.5745  &  0.1022  &  0.5783\\
  April & 0.3221  &  0.0622  &  0.4431  &  0.4106\\
  May & 0.6451  &  0.5700  &  0.0985  &  0.1154\\
  June & 0.5260  &  0.9274  &  0.6738  &  0.7060\\
  July & 0.5384  &  0.5004  &  0.3646  &  0.5631\\
  August & 0.1553  &  0.3217  &  0.0180  &  0.0885\\
  September & 0.1081  &  0.3099  &  0.4733  &  0.3288\\
  October & 0.4983  &  0.8576  &  0.5301  &  0.4271\\
  November & 0.1029  &  0.0137  &  0.1615  &  0.4746\\
  December & 0.7882  &  0.0841  &  0.4347  &  0.0144\\
  \hline
\end{tabular}
\caption{P-values of the portmanteau test defined in eq.~(\ref{QL}) with $L=15$.}
\label{pval_saskat}
\end{table}
}


\subsection{Saugeen river}
The Saugeen River is located in southern Ontario, Canada; it begins in the Osprey Wetland Conservation Lands and flows generally north-west about 160 kilometres (99 miles) before exiting into Lake Huron. Starting from 1950 it is served by Saugeen Valley Conservation Authority (SVCA), a corporate body founded for managing and preserving water and other natural resources in river watershed. It is also among the most important areas in Ontario for fishing, specifically nearby significant dykes as Denny's Dam, in which a popular conservation area is located.

Data analyzed are average monthly flows from January 1915 until December 1976, measured at Walkerton, Ontario.

\begin{figure}[ht]
		\centering{}
		\includegraphics[width=0.95\textwidth]{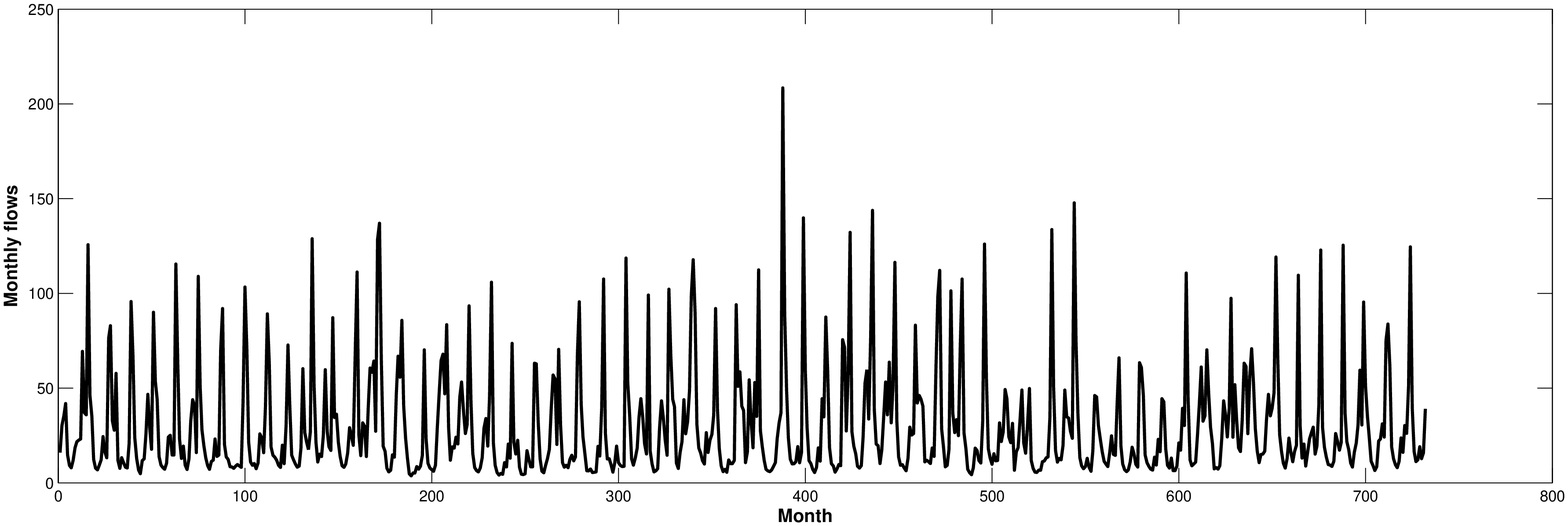}
		\centering{}
		\includegraphics[width=0.95\textwidth]{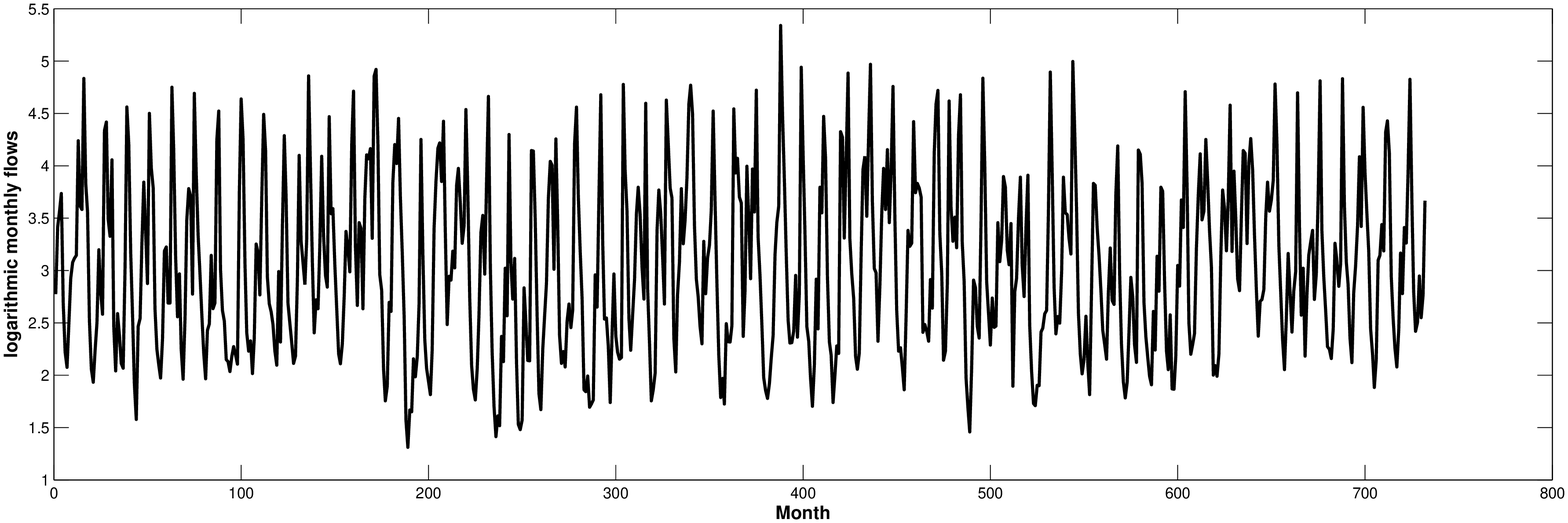}
	\caption{Monthly flows (up) and logarithmic monthly flows for Saugeen River}
	\label{fig:saugeen_plot}
\end{figure}

\begin{table}
	\centering
	\begin{tabular}{|c|c|c|c|c|c|}
	\hline
		Model & Years of changepoint & $RMSE$ & $MAE$ & $MAPE$ & $Fitness$\\
		\hline
		PAR$_{(0;1)}$ & / & 0.4851 & 0.3712 & 11.1849 & 1.1871 \\
		\hline
		PAR$_{(1;1;7)}$ & 1970 & \textbf{0.3520} & \textbf{0.2869} & \textbf{9.0176} & 1.1918 \\
		\hline
		PAR$_{(2;3;5)}$ & 1965, 1970 & 0.3756 & 0.2961 & 9.3380 & 1.1944 \\
		\hline
		PAR$_{(3;2;7)}$ & 1950, 1958, 1970  & 0.3762 & 0.2966 & 9.2641 & 1.1972\\
		\hline
	\end{tabular}
	\caption{Results of evaluation criteria of the logarithmic forecast errors for Saugeen}
	\label{tab:saugeen}
\end{table}

Table \ref{tab:saugeen} shows results of optimal PAR models with $0, 1, 2, 3$   changepoints: year 1970 is always detected as changepoint in our models. One possible reason would be related to works aiming to reconstructing Denny's Dam. In fact, between the end of 1960s and the beginning of 1970s, Great Lakes Fishery Commission managed to rebuild Denny's Dam in order to provide an effective bloackage against parasites such as sea lamprey, preventing them from infiltrating in river. Being Denny's Dam among the biggest dykes of river course this could have been a non-ignorable effect on its flow. There have also been important human work on Saugeen in the 1950s: the main reason of SVCA creation in 1950 was, indeed, flood control management. Walkerton business district, which is the gauging station, has been subject to major floods in early and mid 1900. This has led to the construction, starting from 1956, of 2.4 km of dykes and floodwalls to protect the central business district as well as residential neighborhoods from potential floods.

\begin{figure}
\centering
\includegraphics[width=15cm, height=.25\linewidth]{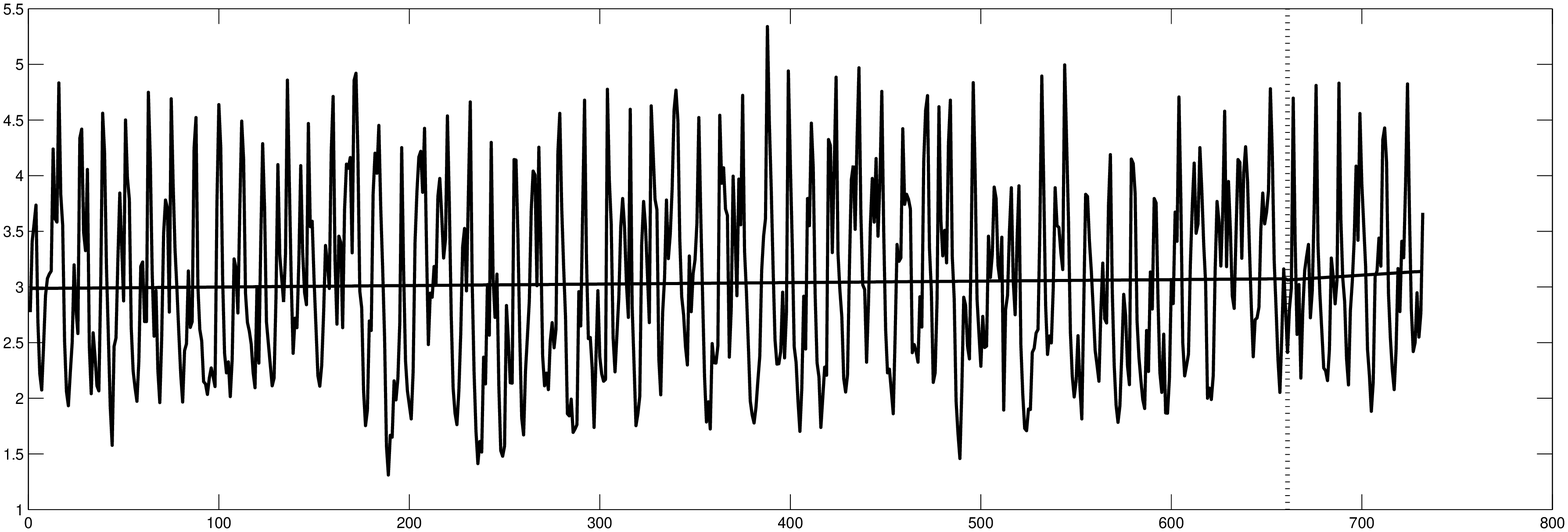}
\caption{Changepoint detected on year 1970 for Saugeen}
\label{fig:saugeen_changepoint}
\end{figure}

As far as goodness of fit is concerned, we observe that also in this case fitness is monotonically increasing with number of changepoints; in terms of forecasting the best performance is observed for model with single changepoint in 1970 considering all measures. Figure~\ref{fig:saugeen_changepoint} plots time series with this changepoint.

As far as this time series has already been modeled in literature we can also perform some comparisons. \citet{WIZX07} proposed a functional-coefficient autoregression model in order to estimate and forecast monthly flows of Saugeen. Forecasting performance, measured on natural data, have been compared to PAR(1) model by \citet{HM94}, resulting in an improvement in terms of $MAE$ from $10.8986$ to $10.3689$. Corresponding value of our model $PAR_{(1; 1; 7)}$ computed on natural data is $9.4827$, which further improves standard PAR(1) model.

\begin{figure}
\centering
\includegraphics[width=\linewidth, height=5cm]{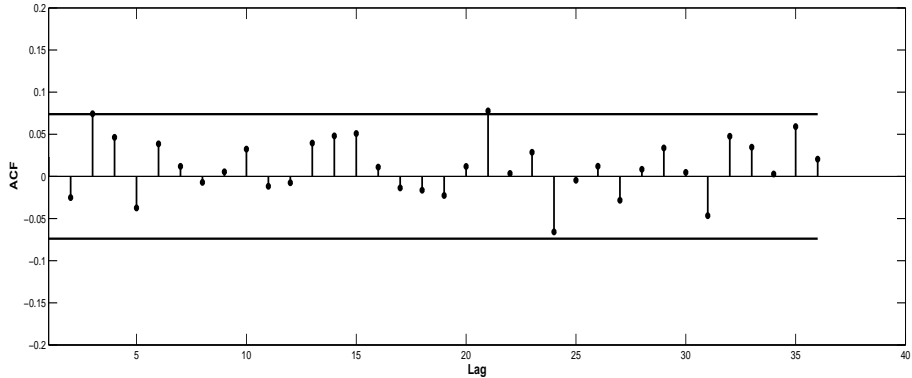}
\caption{Autocorrelation function (ACF) of the residuals of the fitted PAR models with three changepoints to the Saugeen flow.} \label{fig:ACF_log_saugeen}
\end{figure}

Autocorrelation function (ACF) for the residuals from the PAR model with one changepoints is reported in Figure \ref{fig:ACF_log_saugeen} and it shows very few significant autocorrelations. P-values of portmanteau test for the model are reported in Table \ref{pval_garonne}. These values indicate that model is not rejected at $5\%$ significance level, except for September of first segment.

\begin{table}
\centering
\begin{tabular}{|c|c|c|}
  \hline
   & 1915-1969 & 1970-1976 \\
   \hline
  January  & 0.6072 & 0.7949  \\
  February & 0.7640 & 0.2823  \\
  March    & 0.9022 & 0.4626  \\
  April    & 0.5054 & 0.3544  \\
  May      & 0.1796 & 0.2453  \\
  June     &  0.8815 & 0.6919  \\
  July     &  0.4122 & 0.6855  \\
  August   &  0.5044 & 0.2895  \\
  September&  0.0129 & 0.2006  \\
  October  & 0.3257 & 0.3474 \\
  November & 0.8046 & 0.3445  \\
  December & 0.6125 & 0.1522  \\
  \hline
\end{tabular}
\caption{P-values of the portmanteau test defined in eq.~(\ref{QL}) with $L=15$.}
\label{pval_saugeen}
\end{table}

\section{Conclusions}
\label{sec:concl}
The goal of our research was to develop a computational procedure to estimate the number of changepoints and their locations in time series with periodic structure. We analyzed river flow series, for which accurate forecasting is a highly demanding issues in hydrology, especially for the management of reservoir systems. An undetected changepoint can lead to much less accurate forecasting. Hence identifying a changepoint becomes an important task as the rapid environmental change act differently on river flows: the magnitude and timing of river flows change, and new patterns of extreme droughts or floods are observed.

Our procedure estimated changepoints for time series related to river flows of Garonne (France), Saugeen and South Saskatchewan (Canada). The reasons for such changepoints are possibly due to both human activities and climatic oscillations. The comparison of our findings with forecasting results of~\citet{NMH85} and~\citet{WIZX07} confirms efficiency of the proposed method.

In our study, we examined monthly data with changepoints allowed only at the end of the year (that is, a multiple of number of seasons). Modifications of method proposed in the present paper are under study: techniques for monthly, weekly or daily time series with periodic structure allowing changepoints at any season are worth pursuing. Therefore, detecting a changepoint in the middle of a year will prevent dispersing its effects over adjacent seasons. Moreover, as far as PAR models are based on a large number of parameters, one could question on whether it is necessary to consider a separate AR model for each season: we allowed to build subset PAR models in order to conveniently decrease number of parameters, but a considerable gain in parsimony would be achieved by reducing number of seasons in PAR model (\cite{FP04} and \cite{HM94} proposes several statistical hypothesis tests).

Lastly, it is known that a stationary autoregressive processes has a short memory (\cite{BD13,Ro03}). Time series which exhibit long range dependence are characterised by autocorrelations which decay very slowly, while a stationary autoregressive process have rapidly decaying autocorrelations. Hydrological data or internet traffic data generally exhibit structural changes and long range dependence~\citep{SB13}. Therefore a long memory process with periodic structure could be appropriate for hydrological data.

\begin{acknowledgements}
The authors thank professor Francesco Battaglia for his valuable and constructive remarks. Part of this work has been supported by the project Green Water Labex Cote.
\end{acknowledgements}
\appendix

\noindent\textbf{Conflict of Interest Statement\\}
The authors declare that they have no conflict of interest.
\bibliographystyle{spbasic}
\bibliography{changepoint_PAR}


\end{document}